\documentclass[10pt,a4paper,twocolumn]{article}

\date{}

\author{H. GENREITH
\footnote{\em Email:genreith@geo.uni-koeln.de}
\\Institut f\"ur Geophysik und Meteorologie\\ 
der Universit\"at zu K\"oln}

\title{The Large Numbers Hypothesis:
Outline of a self-similar quantum-cosmological Model}

\begin{document}

\maketitle

\begin{abstract}
In 1919 A. Einstein suspected first
that gravitational fields could play an essential role in the
structure of elementary particles.
In 1937, P.A.M. Dirac found a miraculous link
between the properties of the visible Universe and elementary particles.
Both conjectures stayed alive through the following decades but still
no final theory could be derived to this issues.
The herein suggested fractal model of the Universe gives a
consistent explanation
to Dirac's Large Numbers Hypothesis and combines the conjectures of
Einstein and Dirac.
\end{abstract}

\section{Introduction}

\subsection{The Large Numbers Hypothesis}

Dirac\cite{dirac,dirac2} first mentioned in 1937
a presumable connection between the properties of elementary particles
and the Universe by multiplying elementary properties with certain
large numbers
\footnote{Nota bene: The Large Number $\Lambda$ has nothing in common with
the cosmological constant $\Lambda$.}.
Since then this conjecture is called the Large Numbers Hypothesis.
It states, that the cosmological quantities $M,R,T$ can be related to the
particle quantities $m,r,t$ through the scale relation\cite{carneiro1}
\begin{equation} \label{LambdaC}
\frac{T}{t}=\frac{R}{r}=\sqrt{\frac{M}{m}}=\Lambda \sim 10^{38 - 41}
\end{equation}
which may easily be verified for example for the properties
of the Universe and the proton.
Mass, time and length is enough to construct a complete system of units.
A simple dimensional analysis leads to a scaled quantum of action
\begin{equation}
\label{Lambda}
H=\Lambda^3 \cdot h
\end{equation}
in order to handle the Universe like an elementary particle.
Hence the Universe should have an angular momentum of about $H/2\pi$
and it can be
suspected to be a huge rotating black-hole
close to its Kerr limit, or close
to G\"odel's spin \cite{carneiro1,carneiro2,sidharth4,godel},
which is the value for the rotating
cosmological solution to Einstein's equations of General Relativity.

The recent discussion regarding the origin of and
destiny of the Universe raises, for example, the question of where
the Universe comes from. The main explanation
for this is that time and space were created with the big-bang and
therefore the question of where the Universe is
embedded or what happened before the creation,
is not admissible.
This takes the assumption that
there was no reason for the creation and that the Universe was created
from 'nothing' by a type of quantum
fluctuation and should have an overall-energy of zero.
Although this explanation sounds plausible, it does not solve
the real problem that if the Universe was created from 'nothing',
why should there not be countless other Universes,
and what relationship could there be between them?
Numerous scientists have tried to give an explanation to this
question (e.g. A. Linde~\cite{linde1,linde2}).
Whatever the correct answer is, the explanation must be a kind of infinite
regress, as otherwise the embedding problem would recur.

An additional question that is also unsatisfactory
answered is, what matter, such as elementary particles, really is.
To answer this question, one tries to find the most
elementary particles by experimentation and theory, all matter should
be composed of these smallest 'grains' of
matter. But this explanation always raises the question that
these 'grains' could be composed of something that could
be described by an even more elementary description.
This is where the so-called string-theory comes in; this gives a
topological explanation, with the strings as the origin of matter.
But what are these strings composed of, they should
be pure topology, as space-time curvatures are~\cite{einstein,kaku}.

\subsection{Can the Universe be described as a Black Hole ?}

The critical mass density of the Universe is the amount of mass which
is needed to
bring the universal expansion to a halt in the future  and to a collapse
in a final big crunch. This mass is, depending on the world model,
about $10^{53} $ kg.
The visible mass of the Universe
is sufficient for only about some percent of the critical mass density.
On the other hand
the mass density of the baryons
should be 10 to 12 percent of the critical mass density~\cite{coughlan}
as can be derived from the theory of nucleosynthesis and
the measured photon density.
Therefore the so-called dark matter should be about 10 times
larger than the visible bright mass. This matter shows itself in the
extinction of light on its way through the Universe as well as by its
gravitational force on galaxies and galaxy clusters.
Including this, the mass density of baryonic matter seems to be not
large enough to close the Universe.
Besides the classical baryonic matter therefore
also exotic matter is taken into consideration as far as 
the overall mass of the Universe is concerned.

The Schwarz\-schild-radius
$R_{\rm{ss}} = \frac{2GM}{c^2}$ of a given mass $M$ is derived from the
Einstein field equations considering a point-like mass.
The Schwarzschild solution is a static,
homogeneous and isotropic  
solution for the region outside the Schwarzschild radius.
However, the inside solution may look other.
A very interesting solution is the
solution of Oppenheimer and Snyder~\cite{sexl1,oppenheimer} which shows the
astonishing result that the inside solution must
be a Friedmann-Universe \cite{gibbs}.
This result is an outcome from the fitting of the outside
to the inside solution of a
collapsing star: after the burning out of the star every pressure vanishes
when the star shrinks to 9/8-th of its Schwarzschild-radius.
When the collapsing
star reaches the event-horizon one has to set $p=0$ and one gets a
Friedmann-Universe
at its maximum expansion. So the Oppenheimer/Snyder-solution may be
taken for a
speculation of a Friedmann-Universe inside a black hole, shrinking
down to a singularity and from there expanding back
to its maximum expansion again.

If someone does not like such an interpretation, there is the problem to
explain how the expanding Universe let behind its own event-horizon.
As the standard Big-Bang Universe starts to expand
at vanishing spatial dimensions,
this leads to the following seemingly paradoxical context:
Either the Schwarz\-schild-radius is greater-equal to the
world radius
\footnote{Although the
dimension of the world radius and the Schwarzschild-radius may be
equal, there is a considerable  
difference: The world radius is the 'visible' dimension which is
defined by an infinite red shift. 
On the other hand, the  Schwarzschild-radius can never be seen. Even
when an observer is
placed close to this border (as seen from 'outside'): In this case the
line of sight would be  
curved when seen from outside but an inside observer would observe it
as free of any forces and  
straight. Because of this any number of world radius' can
be placed into the area of the Schwarzschild- 
radius even if both have the same dimension.} of the Universe,
as large estimations of the  
Universe mass assume, then the Universe may be called a black hole
\footnote{A black hole, if watched from inside,
may be called a white hole because it is guessed that from its inside
singularity everything can come into existence~\cite{kiefer,gibbs}.
A black hole actually needs an outer space which it is defined on,
but the existence or non-existence of an  
outer space is first of all a question of belief.}.
Or, as small estimates of the mass of the Universe assume,
the Schwarz\-schild-radius is  about  
some 10 percent of the today world radius of the Universe.
But then the Universe should have expanded beyond
\footnote{A 'pushing in front' of the event- 
horizon is easily possible due to the increasing scale-factor $R(t)$
of the Universe.}
its Schwarz\-schild-radius in former times when it
crossed a radius of at least
1.5 billion light-years, which is an event that is supposed to be not in
casual harmony with general relativity.
By this one could guess that our Universe should be indeed a
black hole.

The theory of inflation~\cite{linde2} tries to avoid such a paradox by a
kind of extremely fast inflationary expansion.
But finally the standard inflationary model
also demands a critical mass density for the Universe.
Recent models prefer cosmologies with a cosmological constant
$\Lambda \ne 0$ and small densities, which are also in agreement with the
interpretation of the luminosities of far away Type-Ia-Supernovae
which seem to be about $0.3m$ darker than expected.
These more complex $\Lambda \ne 0$-models open a large variety of possible
large scale structures of the Universe. It shall be emphasized,
that especial the non-critical, so-called open (hyperbolic)
Universes, must not have an infinite volume in any case\cite{parker}.
The correspondance between
an open and an infinite Universe is only true for the very special case of
a zero cosmological constant in a simple-connected topology of the
Universe \cite{roukema}. Also there exist competing theories to explain
the luminosities of the Type-Ia-Supernovas, such as taking into
account the nonhomogenity\cite{sanchez} of the Universe
on medium scales and others. And the work of Gawiser and Silk
 \cite{gawiser}, which numerically calculates the 10 most
discussed cosmological models and
relates them to the variation of the observed cosmic microwave background
(CMB), shows the astonishing result
that the critical standard model fits the observation of the CMB
by far best.

Therefore, as a working-hypothesis,
herein a black-hole-Universe is defined through a Universe of
critical mass density. The possibility of a black-hole Universe was already
taken into consideration in the literature.
Just as examples, out of the uncounted works about this idea, are cited
here the books of
J-P.Luminet~\cite{luminet}and Smolin~\cite{smolin}. In the 
works of Carneiro \cite{carneiro1,carneiro2} the Universe is considered to
be a huge rotating black hole \cite{carneiro1}
with remarkable links between particle physics and cosmology.
A model of a Universe as a white hole is also provided by Gibbs \cite{gibbs}.

\section {Simple solutions for a black-hole-Universe}

The main sections of the model\cite{genreith}
are repeated\footnote{You may skip to section \ref{tmotn}.}
here for convenience.
The coordinate frame used herein is, if not mentioned differently,
for an observer at spatial infinity.

\subsection{The Einstein-deSitter model}

The Einstein-deSitter model (EdS model) follows from the Friedmann-models
which are simplified with a curvature parameter $\kappa=0$ and
a cosmological constant $\Lambda=0$.
Therefore the Universe is approximated with an euclidic,
isotropic and homogeneous world model.
Hence the scale-factor  $R(t)$  is a solution of
$R \dot R^2 = \textrm{Const.}$, which gives
\begin{equation}
R(t_{\rm e}) = R_{\rm 0}\cdot { (\frac{t_{\rm e}}{t_{\rm 0}}) }^{2/3}
\quad \rm{.}
\label{EdSSkala}
\end{equation}
Therein $t_{\rm{e}}$ is the time of emission of a signal 
and $t_{\rm{e}}=0$ is the time of the big bang and 
$t_{\rm 0}$ is the present time. 
With the Hubble equation $\frac{dR}{dt} = H(t) R(t)$ follow
the relations for the age and size of the Universe:
\begin{displaymath}
t_{\rm{U}} = \frac {2} {3H_{\rm 0}}
\quad \rm{and} \quad
R_{\rm{U}} = \frac {2c} {H_{\rm 0}}
\end{displaymath}
The actual distance $r$ between two points with previous distance $\rho$
is derived from the equation:
\begin{displaymath} r(t) = R(t) \cdot \rho \end{displaymath}
The standardization is given by reference to the present time
$ R(t_{\rm 0}) = R_{\rm 0} =: 1 $~.

General Relativity demands a maximal velocity  $c$ only for the
peculiar movement.
The variation of the scale-factor
\begin{displaymath}
\frac{dR}{dt_{\rm e}}
= \frac{2 R_{\rm 0}}{3  t_{\rm 0}^{2/3} t_{\rm e}^{1/3}}
\end{displaymath}
runs to infinity for small times of emission.
Therefore the overall-velocity of the Expansion
\begin{equation}
\frac {dr} {dt} = \frac {dR} {dt}
\cdot \rho + R \cdot \frac { d \rho } {dt}
\label{drdt}
\end{equation}
can be much greater than the speed of light.
Hence, considering events for which the speed of light is a given limit,
one has to look at variations of $\rho$.

The fact that the EdS-model is simplified with a curvature
parameter $\kappa=0$
seems to point out (as (\ref{drdt}) never turns its sign to minus)
that it is not suitable to  
treat a black-hole Universe which should have an $\kappa=+1$ and should
collapse in a finite  
time. The critical EdS-Universe is, in the scope of this simple model,
a black-hole-Universe which collapses in an infinite time.

\subsection{The Planck dimensions} \label{diePldims}

If one equates the Planck energy $E = \hbar\omega $
with the  Einstein energy $ E = m_{\rm 0}c^2$ of
a mass charged particle
one gets the de-Broglie wavelength of a resting particle $m_{\rm 0}$,
which is also known from the theory of photon scattering on electrons
as the Compton wavelength  
$\lambda_{\rm C}=\frac{\hbar}{m_{\rm 0}c}$.
The wavelength of the particle is in inverse proportion to its mass
than the Schwarz\-schild-radius of the corresponding black-hole
$\rho_{\rm{ss}}=\frac{2Gm}{c^2}$.
The identity of both lengths leads to the Planck dimensions
\footnote{The usual description is $m_{\rm{pl}}
=  \sqrt{ \frac{c\hbar}{G} }$ which follows from just  
dimensional arguments. I will use the definition which follows
from the equality of event-radius to  
Compton-wavelength. See also footnote following equation (\ref{spin}).}:
\begin{eqnarray}
m_{\rm{pl}}  & = &  \sqrt{ \frac{c\hbar}{2G} }
=	1.54\cdot10^{-8} \rm{kg}	\nonumber\\
l_{\rm{pl}}  & = &	\frac{  \hbar  }{ m_{\rm{pl}}  c }
=	\sqrt{ \frac{2 \hbar G }{c^3} }
=	2.29\cdot10^{-35} \rm{m}	\nonumber\\
t_{\rm{pl}}  & = &	\frac{ \hbar }{  m_{\rm{pl}}c^2  }
=	\sqrt{ \frac{2 \hbar G }{c^5}  }
=	0.762\cdot10^{-43} \rm{sec}  \nonumber\\
E_{\rm{pl}}  & = &  m_{\rm{pl}}c^2
=  \sqrt{ \frac{c^5\hbar}{2G} }
= 1.38\cdot10^{9}  \rm{J}  
\label{Pldim} \end{eqnarray}
The main property of the Planck dimension is the fact, that the
energy of a wave with wavelength  
$l_{\rm{pl}}$ equals to a mass which bends space  to a black hole of
Planck size.

\subsection{The mass formula for the Einstein-de Sitter model}

The Eds model delivers the coordinate distance of an event travelling
with the speed of light  
from (\ref{drdt}) through the integration of
$ d\rho = \frac {c \cdot dt} {R(t)} $:
\begin{equation}
\rho=\frac{3ct_{\rm 0}}{R_{\rm 0}}\cdot
\left[ 1-  (\frac{t_{\rm e}}{t_{\rm 0}})^{1/3}
\right]
\label{pkentf}
\end{equation}
This distance should at maximum be equal
to any given Schwarz\-schild-radius
$\rho_{\rm{SS}}=2GM/c^2$ which results in:
\begin{equation}
M(t_{\rm e})= \frac{c^3} {H_{\rm 0}G} \cdot
\left[  1 - \left(\frac{3 H_{\rm 0}t_{\rm e}}{2}\right)^{\frac{1}{3}}
\right]
\label{MF}
\end{equation}
This time-dependent mass should be at least included by a
gravitational spherewave  
starting at time $t_{\rm e}$ and running with velocity $c$.
Otherwise the wave would have to go  
beyond  its own event-horizon.
Inserting $t_{\rm e} = 0$ for the origin of the Universe one gets:
\begin{equation}
M_{\rm{Umin}} \ge M(0)
= \frac{c^3}{H_{\rm 0}G}
\label{MU}
\end{equation}
This is the mass which makes an EdS Universe critical.
The value of $H_{\rm 0}$ is still controversial and differs depending
on the source between  
approximately $50$ and $100 \frac{\rm{km}}{\rm{sec} \rm{Mpc}}$.
So the mass of the Universe should be in the bounds
$M_{\rm U} \in  [1.248,2.497] \cdot10^{53} \rm{kg}$~.
For any computational purpose herein the average value of 
$75 \frac{\rm{km}}{\rm{sec} \rm{Mpc}}$
is used, which agrees also with recently elaborated values 
$(72 \pm 6) \rm{km} /  (\rm{sec} \rm{Mpc}) $ \cite{richtler}
of the Hubble constant.

The mass formula (\ref{MF}) is now generalized to local gravitational
waves running in a local
flat space-time. For that purpose (\ref{MF}) is expanded to a Taylor
series at the time
$t_{\rm 0}$ transforming the time coordinate to $t=t_{\rm 0}-t_{\rm e}$. 
By this one gets the mass formula for small masses implying
small times $t<<t_{\rm{0}}$:
\begin{eqnarray}
m(t) =  \frac{1}{2} \frac{c^3}{G} t
+ \frac{1}{6} \frac{c^3}{G}  \frac{t}{t_{\rm 0}} \cdot t
+ \frac{5}{54} \frac{c^3}{G} \left( \frac{t}{t_{\rm 0}} \right)^2 \cdot t
+ \dots
\label{kleinmt}
\end{eqnarray}
The factor
$ \left(\frac{t}{t_{\rm 0}}\right)^{n}$
rapidly drops to zero for small times and therefore one can calculate
further on with only the first part of the sum:
\begin{equation}
m(t) = \frac{1}{2}\frac{c^3}{G}\cdot t
\label{kleinm}
\end{equation}
For the peculiar distance of events which
travel with the speed of light
one gets similar to the derivation of (\ref{kleinmt})~:
\begin{equation}
\rho (t) =  c t  +  \frac{c H_{\rm 0}}{2} t^2
+  \frac{5}{12} c H_{\rm 0}^2 t^3  + \dots
\end{equation}
Hence the apparent force
\footnote{This force is seen only by an observer at spatial infinity.
An observer travelling inside a black hole would never feel to
hit against the event-horizon because any forced curvature of his
line of sight would be sensed to be straight.}
acting on a gravitational event running with $c$ is:
\begin{equation}
\mid \vec F_{\rm{e}} \mid
= \frac{d}{dt}(mv)
= \frac{c^4}{2G} (1 + 3H_{\rm 0} t + \dots ) \cong \frac{c^4}{2G}
\label{Fe}
\end{equation}
This almost constant force acts on the event over the area
of the Schwarz\-schild-radius:
\begin{equation}
E \approx F_{\rm{e}}\cdot \rho_{\rm{ss}}
= \frac{c^4}{2G} \cdot \frac{2Gm}{c^2}  =  m c^2
\label{Emc}
\end{equation}
It seems that a gravitational wave may run unhindered just if the
mass included in its sphere is zero or infinite.
Every distortion
\footnote{A distortion of such a kind is given in
general by the deviation of the
space-time-curvature through the gravitational wave itself if the
energy of the wave is not
neglectible.}
of space-time causing a mass creates an event-horizon proportional
to this mass. Then the wave hits its self-made horizon.
The rest energy of the so created (virtual) mass is borrowed from
the energy of the gravitational wave and is transformed into a potential
energy of  the same
quantity represented by a virtual-black-hole~\footnote{A 'virtual'
black hole differs from a
'real' black hole by the fact that the energy for its creation does
not come from a fatal
gravitational collapse and by this leaving behind a large potential
well but was borrowed from
the energy of a gravitational wave.}.

\section{An approximate stationary solution for black hole particles}
\label{stNSL}

As a simple approximation one can consider the effect of a wave in a
rectangular potential well.
The gravitational wave runs unhindered in a small area until it is
stopped, as mentioned from
outside, by its self-made event-horizon which exerts a nearly  infinite
apparent force
$F_{\rm{e}}$ which hinders the progress of the wave
\footnote{One may also imagine this
wave as a wave travelling with $c$ along the event-horizon.}~:
\begin{eqnarray}
V & = &0  \qquad  \,\, \textrm{for} \quad r \in [0,\rho_{\rm{SS}}]
\nopagebreak[3]
\nonumber\\
V & = & \infty \qquad  \textrm{for} \quad r >  \rho_{\rm{SS}}
\label{Pot0}
\end{eqnarray}
The common known solution of the time-in\-de\-pen\-dent
Schr\"o\-din\-ger equa\-tion for a
particle in such a rectangular potential well delivers the
energy eigenvalues
\begin{equation}
E_{\rm{n}}=\frac{n^2\pi^2 \hbar^2}{2ma^2}
\end{equation}
with $n=1,2,3\dots$ and  $a$ the diameter of the well.
With the substitution of $a = 2\rho_{\rm{SS}}= \frac{4Gm}{c^2}$  and
$m = \frac{E_{\rm{n}}}{c^2}$
one gets the  energy eigenvalues as the real roots of
$ E_{\rm{n}}^4=\frac{n^2\pi^2 \hbar^2 c^{10}}{32 G^2} $~:
\begin{equation}
E_{\rm{n}} = \pm \frac{\sqrt{\pi}}{2^{3/4}}  \sqrt{n} E_{\rm{pl}}
\end{equation}

The factor $ \frac{\sqrt{\pi}}{2^{3/4}}  = 1.054$ origins from
the simplification of the
potential (\ref{Pot0}) and is set equal to 1.
Then the masses of the virtual Schwarz\-schild areas have the
eigenvalues
\footnote{The same quantization rule for black holes was found
for example by Khriplovich~\cite{khriplovich} from a totally different
point of view.
For the handling of negative masses see also the article
of Olavo \cite{olavo}.}~:
\begin{equation}
m^{\pm}_{\rm n}=\pm \sqrt{n}\cdot m_{\rm{pl}}
\label{mn}
\end{equation}
The difference of neighbouring positive eigenvalues is 
\begin{equation}
\Delta m_{\rm n} = ( \sqrt{n+1} - \sqrt{n} ) m_{\rm{pl}}
\cong \frac{1}{2\sqrt{n}}\cdot m_{\rm{pl}}
\label{dmn}
\end{equation}
for large $n$. From the equations (\ref{mn}) and (\ref{dmn}) follows
\begin{equation}
\Delta m_{\rm n} = \frac{1}{2} \cdot \frac{m_{\rm{pl}}^2}{m_{\rm n}}
\qquad \textrm{with} \qquad
n= \frac{1}{4} \cdot \frac{m_{\rm{pl}}^2}{\Delta m_{\rm n}^2}
\label{dmn1}
\end{equation}
and the relation between stimulated mass $\Delta m_{\rm n}$ 
and  positive virtual mass $m_{\rm n}$ is:
\begin{equation}
\label{twon}
\frac{\Delta m_{\rm n}}{m_{\rm n}} =  \frac{1}{2n}
\end{equation}
The radius of an elementary particle in this model is the
event-radius of the virtual mass
\begin{equation}
\rho_{\rm n} := \rho_{\rm{SS}}(m_{\rm n}) = \frac{2Gm_{\rm n}}{c^2}
\end{equation}

The formula (\ref{dmn1}) therefore serves the equation
$\lambda_{\rm C} = \frac{\hbar}{m_{\rm 0}c}
\qquad \Leftrightarrow \qquad
m_{\rm 0}\cdot \lambda_{\rm C} = \frac{\hbar}{c}$
for the Compton wavelength of an elementary particle:
\begin{equation}
\Delta m_{\rm n} \cdot 2 \rho_{\rm n}
=  \frac{m_{\rm{pl}}^2}{2m_{\rm n}} \cdot \frac{4Gm_{\rm n}}{c^2}
=  \frac{\hbar}{c}
\end{equation}
with the substitutions $m_{\rm 0}=\Delta m_{\rm n}$
and $\lambda_{\rm C} = 2\rho_{\rm n}$.

Heisenberg's uncertainty relation is fulfilled,
during the creation of the virtual black hole, 
for the stimulated mass:
\begin{equation}
\Delta E \Delta t
= \Delta E_{\rm{n}} \cdot \frac{\Delta x}{c}
= \Delta m_{\rm n}c^2 \cdot \frac{2Gm_{\rm n}}{c^2\cdot c}
= \frac{Gm_{\rm{pl}}^2 }{c} = \frac{\hbar}{2}
\end{equation}

The angular momentum $ \vec{L}=m\vec{v} \times \vec{r} $
of a wave with mass $\Delta m_{\rm n}$
which circulates at the speed of light $c$
along the horizon of a black hole with mass $m_{\rm n}$ is:
\begin{eqnarray}
\mid \vec{L} \mid & = &\Delta m_{\rm n} c \cdot \rho_{\rm n}
\nonumber\\
& = & \frac{m_{\rm{pl}}}{2\sqrt{n}} c \cdot
\frac{2G}{c^2}\sqrt{n}m_{\rm{pl}} =
\frac{G}{c}\cdot\frac{c\hbar}{2G}=\frac{\hbar}{2}
\nonumber\\
\Rightarrow s_{\rm{z}}  & = &  \pm \,  \frac{\hbar}{2}
\label{spin}
\end{eqnarray}
The spin of a stimulated black hole particle is, by this, half of
Planck's quantum, which corresponds to a fermion
\footnote{However the value of the spin
$ \mid \vec{L} \mid = \frac{G}{c} m_{\rm{pl}}^2 $
relates to the definition of the Planck mass:
If one equates two instead of one Compton wavelength
to the Schwarz\-schild-radius
one gets $\sqrt{\frac{c\hbar}{G}}$ for the Planckmass and
a spin of $\hbar$, which corresponds to a boson.}.
Particles relating to the energy of stimulation of 
a virtual-miniature-black-hole (\ref{dmn}) will herein be refered
to as SBH-particles.


\begin{table*}

\caption{Chart of SBH-particles:
The entries in italics for the  'SBH-Massless'
$(n\rightarrow\infty)$ and the
'SBH-Universe' $(n \rightarrow 0)$ are values introduced by an
extrapolation of the formulas
of chapter \ref{stNSL} to their limits.
The value for the
SBH-massless arises if one identifies the virtual mass $m_{\rm n}$
with the mass of the Universe from equation (\ref{MU}),
and the value of the mass of the SBH-Universe arises if one identifies
the stimulated mass $ \Delta m_{\rm n}$ with the mass of the Universe.
The mass of a quark is estimated here as $\sim 10$ MeV. }

\label{tab}
\[\begin{tabular}{lllllll}
\hline
Eigenvalue & virtual mass & stim. mass &
& SS-diam. & Hawk. time & Name
\\
$ n $ & $m_{\rm{n}}$  & $\Delta m_{\rm{n}} $  & $\Delta E_{\rm{n}} $
& $2\rho_{\rm{n}}=\lambda_{\rm{C}}$ & $\sim t_{\rm{V}} $
& of stim. mass
\\
$ [1] $ & $[\rm{kg}]$ & $[\rm{kg}]$ & $[\rm{eV}]$ & $ [\rm{m}] $
& $ [\rm{sec}] $  & $m_{\rm 0}=\Delta m_{\rm n}$
\\\hline &&&&&&\\
$ \sim 0 $ & $ \pm 7.1\cdot10^{-70} $
& $ 1.7\cdot10^{53} $ & $9.3\cdot10^{88}$
& $  2.1\cdot10^{-96} $ & $ \approx 0
$ & $ \textit{SBH-Universe} $
\\
$  1 $ & $ \pm 1.5\cdot10^{-8} $ & $ 7.7\cdot10^{-9}$ & $0.4\cdot10^{28}$
& $ 4.6\cdot10^{-35} $ & $ \sim10^{-43} $ & Planck quant
\\ &&&&&&
\\
$  2.1\cdot10^{37} $ & $ \pm 7.1\cdot10^{10} $
& $ 1.7\cdot10^{-27}$ & $938\,\rm{[MeV]}$
& $ 2.1\cdot10^{-16} $ & $ \sim10^{12} $ &  SBH-proton
\\
$ 1.9\cdot10^{41} $ & $ \pm 6.6\cdot10^{12} $
& $  1.8\cdot10^{-29}$ & $10\,\rm{[MeV]}$
& $ 2.0\cdot10^{-14} $ & $ \sim10^{18} $ &  SBH-quark
\\
$ 7.2\cdot10^{43} $ & $ \pm 1.3\cdot10^{14} $
& $ 9.1\cdot10^{-31}$ & $0.51\,\rm{[MeV]}$
& $ 3.9\cdot10^{-13} $ & $ \sim10^{22} $ &  SBH-elektron
\\ &&&&&&
\\
$ 1.2\cdot10^{122} $ & $ \pm 1.7\cdot10^{53} $
& $ 7.1\cdot10^{-70}$ & $0.40\cdot10^{-33}$
& $ 5.0\cdot10^{26} $ & $  \sim10^{139} $ &
$ \textit{SBH-massless} $
\\
\hline
\end{tabular}
\]
\end{table*}

\subsection{SBH-Particles}

The table (\ref{tab}) shows the values of some selected masses refering
to the formulas of chapter \ref{stNSL} and gives an idea of an
Universe which is built up in a fractal manner of black-hole-topologies.
All masses origin from stimulated curvatures of space-time
and every black hole has its typical quantum.
For macroscopic black holes these quantums have
energies much smaller than
the electron rest mass in the area of (practically) massless
\footnote{A massless particle in classical quantum dynamics may also
have at least a very small rest mass $\approx 0$ according to
the uncertainty relation
$\Delta E \Delta t = \Delta E \cdot t_{\rm 0} \ge \hbar/2
\Leftrightarrow
\Delta E \ge \hbar/(2t_{\rm 0})
= \frac{3}{4} \hbar H_{\rm 0} \approx 10^{-33} \rm{eV}
$ if one considers a particle blurred\cite{parker} over the whole Universe.}
particles like neutrinos or photons:
a macroscopic black hole seems to radiate thermally like a black body.
But for virtual-miniature-black-holes the stimulated energies are
in the typical bounds of the
well known  elementary particle restmasses, matching their typical
properties as rest energy, Compton wavelength and spin. 

The Hawking-times $t_{\rm V}$ are a rough hint for the lifetimes of
SBH-particles and they show
meaningful values for stable elementary particles.
But effects 
of quantum gravitation should generate very different values for
this lifetimes, namely those
experimentally seen values which can be much less for instable and
much more for stable particles.

The product of virtual mass and stimulated mass is always a
constant for every SBH-particle:
\begin{eqnarray}
C^{\pm}_{\rm{m_{\rm n}}} &:=  \Delta m_{\rm n} \cdot m^{\pm}_{\rm n}
= \pm \frac{1}{2} m_{\rm{pl}}^2
= \pm \frac{c\hbar}{4G}        \nonumber
\\ &=\pm 1.185\cdot10^{-16} \, \rm{kg}^2
\label{Cmn}
\end{eqnarray}

The basic eigenvalue $n=1$ of self-curvature is given by the
Planck mass which has a stimulated mass of approximately the same quantity.
The sizes of the self-curvatures of space-time increase with $n$ and they
reach their maximum at
$\sqrt{n}=10^{61}$ with the Universe as the largest Schwarz\-schild area.
The stimulated mass
of this SBH-Universe is a rather massless particle,
as are neutrino or photon.
An extrapolation of the $\Delta m_{\rm n}/m_{\rm n}$ dependence for an
elementary SBH-particle to the (not allowed) eigenvalue $n=0$ gives
cause to a speculation of
a nearly massless particle of which the stimulated mass is a Universe
with an incredibly small
lifetime satisfieing the uncertainty relation.
As follows from chapter \ref{stNSL} these quantums may be fermions;
the one with the lightest weight is a
neutrino and the one with the heaviest weight is the Planck quantum.
The speculative extrapolation to the
eigenvalue $n=0$ gives as the heaviest fermion a Universe.
Hence a possible
fractal construction of the world is conceivable,
based on black-areas of different sizes.

\subsection{The entropy of black holes} \label{entropy}

A fundamental question in gravity physics is to explain the high
entropy of black holes.
The Bekenstein-Hawking-entropy~\cite{horowitz} is
$S_{\rm{BH}}=\frac{A}{4G\hbar}$.
The surface of the black hole $A=4\pi\rho_{\rm{n}}^2$
can be deployed
\begin{eqnarray}
S_{\rm{BH}}=\frac{2\pi}{c^3}\cdot n
\end{eqnarray}
and relates the black hole entropy directly to its excitation level $n$.

If one relates the rest-energy of a particle, using the herein derived
mass-quantization
(\ref{mn})(\ref{dmn}), to the Hawking-temperature of a black-hole,
which is
$T_{\rm{H}}=\frac{\hbar c^3}{8\pi k_{\rm{B}}G M }$ , one gets:
\begin{equation}
\frac{E_{\rm{0}}}{T_{\rm{H}}}
=\frac{\Delta m_{\rm{n}} c^2}{T_{\rm{H}}(m_{\rm{n}})} = 2\pi k_{\rm{B}}
\end{equation} 
This shows that the SBH-particle is in thermodynamical
equilibrium \cite{kiefer} with the Hawking-
temperature of the stimulated-black-hole.

\subsection{The mass of the proton}

As pointed out in \cite{genreith}, the gravitational power
of the evolving Universe is greater than the power
needed to create virtual black holes for the first primordial
\begin{equation}
t_{\rm B} = (\frac{28\cdot2^{1/3}}{3})^{3/10}
\Bigl[ \left( \frac{G\hbar}{c^5}  \right)^7
\cdot \frac{1}{H_{\rm 0}^6} \Bigr]^{\frac{1}{20}}
\cong 2\cdot10^{-25}\,\rm{sec}
\label{tB}
\end{equation}
This time may be interpreted as a kind of phase change as the Universe
stops boiling.
On the other hand the uncertainty equation for the Universe
gives at this time a
mass-equivalent of $m_{l}=\frac{\hbar}{2c^2t_B}$.
Inserting (\ref{tB}) results in
\begin{equation}
m_l=0.239\cdot(\frac{\hbar^{13}H_0^6}{c^5G^7})^{\frac{1}{20}}
\cong1.7\cdot m_p
\end{equation}
which is a limiting upper value for SBH-masses
close to the mass of the proton.

\subsection{The mass of the neutrino and the rotation of the Universe}
\label{tmotn}

In this model the neutrino is the less weighted fermion,
which results from a fermionic excitation
of the Universe as the underlying virtual-mass-particle.
The mass relation (\ref{dmn1})
$\Delta m_{\rm n} = \frac{1}{2} \cdot \frac{m_{\rm{pl}}^2}{m_{\rm n}}$
gives a minimum mass for the lightest neutrino of
\begin{equation}
m_{\nu}\geq\frac{\hbar H_{\rm{0}}}{4c^2}
=0.40\cdot10^{-33}\rm{eV}
\end{equation}
which can also be interpreted as
the minimum mass for a particle which is blurred over the whole Universe
as shown by Heisenbergs uncertainty relation.
The relation (\ref{dmn1}) shows resemblance to the Dirac-mass-
term \cite{senjanovic},
which follows from the so-called see-saw-mechanism in GUT-theory:
\begin{equation}
m_{\rm\nu} \simeq \frac{m_{\rm{D}}^2}{M_{\rm R}}
\nonumber
\end{equation}
In this case, the Dirac-mass $M_{\rm{D}}$ could be identified
with the Planckmass $m_{\rm{pl}}$
and the mass of the right-handed neutrino $M_{\rm{R}}$ with the
mass of the Universe.

If the Universe is assumed to be a kind of a fermion
and the SBH-partner of the neutrino, it should have an
intrinsic rotation of about
\begin{eqnarray}
\label{Uspin}
\frac{\hbar}{2}=m_{\rm{\nu}}\omega_{\rm{U}}  R^2_{\rm{U}}
\Rightarrow \omega_{\rm{U}}  = H_{\rm{0}} / 2 
\end{eqnarray}
by substituting the herein derived mass of the neutrino and the
EdS radius and mass of the Universe.
As this is just a rough estimation,
one should expect a rotational ratio of the order of some $H_{\rm{0}}$:
\begin{equation}
\omega_{\rm{U}} \sim H_{\rm{0}}
\end{equation}
Past works \cite{carneiro1,carneiro2,kuhne,nodland,obukhov}
about the claimed observation of a rotation of the Universe
give estimates \cite{kuhne}
of $\omega = (6.5 \pm 0.5) H_{\rm{0}}$.

From equation (\ref{Uspin}) follows the scaled quantum
of action $H$ for the EdS-Universe as
\begin{eqnarray}
\label{Uspin1}
\frac{H}{2}=M_{\rm{U}}\omega_{\rm{U}}  R^2_{\rm{U}}
\Rightarrow H=\frac{4c^5}{GH_{\rm{0}}^2}
\approx 2.5\cdot 10^{88}\frac{\rm{kg}\rm{m}^2}{\rm{sec}}
\end{eqnarray}
and with equation (\ref{Lambda}) the scale-factor
of the Large Numbers Hypothesis can be expressed as:
\begin{equation}
\label{LambdaH}
\Lambda=\sqrt[3]{\frac{H}{h}}
=\left(\frac{4c^5}{h GH_{\rm{0}}^2}\right)^{\frac{1}{3}}
\cong (t_{\rm{pl}}H_{\rm{0}})^{-\frac{2}{3}} = 3.3\cdot 10^{40}
\end{equation}

\subsection{The Large Numbers coincidences}

We now can combine the formulas (\ref{LambdaC}),(\ref{MU}),(\ref{LambdaH}) to
derive the masses of usual matter:
\begin{equation} \label{Lmass}
m = \frac{M} {\Lambda^2} = \frac{c^3}{H_{\rm 0}G \Lambda^2}
\end{equation}
which gives:
\begin{equation}
m = \left(
\frac{\pi^2 \hbar^2 H_{\rm 0} } {4 G c } \right)^{1/3}
\end{equation}
This formula is equivalent to the empirical Weinberg formula
\cite{weinberg,sidharth4} for the mass of the pion
$ m_{\pi} \simeq \left(\frac{\hbar^2 H_{\rm 0} } {G c } \right)^{1/3}$.
The Large Numbers coincidence is the fact, that all elementary particles
are close together considering the large number $\Lambda$:
\begin{equation}
m = \frac{c^3}{H_{\rm 0}G \Lambda^x}
\quad \Rightarrow \quad
x=\frac{\ln{(c^3/(m H_{\rm 0}G )) } }{ \ln{\Lambda} }
\end{equation}
Setting in the masses of the proton, pion and electron,
x is clustered
\footnote{Further analysis shows that massless particles cluster at
$x\simeq 3$, elementary particles at $x \simeq 2$,
Life (cells \dots biotope \dots solar-system) around $x \sim 1$, Galaxies,Clusters,Walls around
$x\sim 1/2$ and the Universe itself is $x=0$. See also \cite{carneiro1}.}
at values close to 2,
which are 1.97411, 1.99458 and 2.05465 respectively.
So all elementary particles are determined by the same~$\Lambda$.

The mass (\ref{Lmass}) can be compared with
the herein derived mass relation (\ref{twon})
$\frac{\Delta m_{\rm n}}{m_{\rm n}} =  \frac{1}{2n}$.
The Compton-radius of the mass is the Schwarzschild radius
$r=2G m_{\rm n}/c^2$ of the related virtual-mass,
which gives $\Delta m_{\rm n}=m=rc^2/4nG$.
Comparing the values for m gives
$r/4n = c/H_{\rm{0}} \Lambda^2$,
and as will be shown later $\Lambda=2n$, this results in:
\begin{equation}
r=\frac{2c}{H_0\Lambda}=\frac{R}{\Lambda}
\end{equation}
So the classical Dirac conjecture (\ref{LambdaC}) is recovered.

\subsection{The electric charge}

Moreover a black hole has not to gravitate because as an effect
of its event-horizon no gravitational waves or gravitons may leave it.
An ordinary black hole gravitates because it
leaves behind the gravitational field of a collapsing object.
If for example a Daemon would cut out a stellar black hole exactly at the
Schwarzschild-border and place
it somewhere else in the Universe, such a  black hole would not
gravitate except by the small mass equivalent of its Hawking radiation.
This mechanism may define elementary particles as the Hawking
radiation of virtual-miniature-black-holes.

By this assumption the Compton-wavelength of a particle
is assoziated with a virtual black hole, 
which means a quantum space-time topology with
at least one curvature according to this size.
Such a virtual-miniature-black-hole is charged with the
constant $C^{\pm}_{\rm{m_{\rm n}}}$ (\ref{Cmn}).
This charge is independent
of the mass of the virtual black hole and could give rise to an
electrical charge by quantum gravity effects: the gravitational force
between a non-gravitating virtual-black-hole and a gravitating
stimulated-mass (\ref{twon}) would be $2n$-times stronger
than the gravitational force between two stimulated masses.

The classical ratio between the gravitational force
$F_{\rm{G}}=G\frac{m^2_{e}}{r^2}$ and the
electromagnetic force
$F_{\rm{Q}}=\frac{1}{4\pi\varepsilon_{\rm{0}}}\cdot\frac{e^2}{r^2}$
for for example an electron in any given distance $r$ is:
\begin{equation}
\label{fqfg}
\frac{F_{\rm{Q}}}{F_{\rm{G}}}
=\frac{e^2}{4\pi \varepsilon_{\rm{0}} G m^2_{\rm{e}} }
=4.167\cdot 10^{42}
\end{equation}
This classical ratio
is also a worth mentioning Large Number and it
lies just between the eigenvalues for SBH-quarks and SBH-electrons.
If the virtual mass of a SBH-particle would be present as a
higher order effect
in any unknown way, the gravitational force between a SBH-particle
and the virtual-SBH-mass of
its vis-a-vis would be: 
\begin{equation}
F_{\rm{G-virtual}}=G\frac{\Delta m_{\rm{n}}\cdot m_{\rm{n}}}{r^2}
= \frac{c\hbar}{4r^2}
\end{equation}
As this force is independent of $n$ and therefore independent of
the mass of the particle
it can be related to an electromagnetic charge:
\begin{equation}
F_{\rm{Q}}=\frac{1}{4\pi\varepsilon_{\rm{0}}}\cdot\frac{Q^2}{r^2}
=F_{\rm{G-virtual}}=  
\frac{c\hbar}{4r^2}
\end{equation}
which gives
\begin{equation}
Q=\pm \sqrt{\pi\varepsilon_{\rm{0}}c\hbar}=\pm 5.853\cdot e
\end{equation}
and is about 6 times the elementary charge. As this assumption is
just a plain one (as it must be an effect of higher order), this
charge is not so far away from unity as it seems at first sight.
Hence the Large Number (\ref{fqfg}) represents the ratio of
virtual mass to particle mass:
\begin{equation}
\Lambda \simeq
\frac{ F_{\rm{G-virtual}} }{ F_{\rm{G}} } =
\frac{ m_{\rm{n}} }{  \Delta m_{\rm{n}} }=2n \sim 10^{38 - 44}
\end{equation}

\subsection{On the quantum-topology of the SBH}

To considere elementary particle as a static sphere may be to much
simplified. Especially one could expect a rotating black hole
close to its Kerr-limit,
which gives a more complified model
\cite{sidharth1,sidharth2,sidharth3}.
The equivalence principle of GR provides the
assumption that matter should be equal to a related space-time curvature.
Hence the quantum-topology of a SBH may be for example a torus, with
one main-curvature related to the Planck-length and the other
main-curvature related to the Compton-length. Then it would appear
to be a closed string. Another consideration could be to imagine
the hadrons as a bag model\cite{lee,schvellinger}
with bumps in it, so that the overall-curvature
might be of Compton-size and the bumps providing curvatures which
relates to the quark masses and fractional charges.

It was A. Einstein\cite{einstein} himself who first suspected elementary
particles to be build up by gravitational fields.
An attempt in recent time was done by Recami et al\cite{recami},
describing hadrons as strong-black-holes by a concept of strong-gravity.
The Recami-model describes hadrons and
its constitutents with the, according to the large numbers hypothesis,
scaled Einstein-field equations of GR. This bi-scale theory of gravitational
and strong interactions delivers a description of the confinement and
asymptotic freedom of the quarks, the Yukawa-potential of strong
interactions and derives a strong coupling constant and a mesonic
mass spectra, which fit well the theoretical and experimental values.

A crucial requirement for SBH-particles is that this particles
should behave as known
particles do. So what happens if two SBH-particles collide ?
SBH-particles have at least
two quantum-numbers in this plain model:
The spin $s_{\rm{z}}=\pm \hbar/2$ and the
virtual-mass-charge
$C^{\pm}_{\rm{m_{\rm n}}}=\pm c\hbar/4G$, which corresponds to the
electrical charge $\pm e$ and a virtual-mass of $\pm m_{\rm n}$. 
A SBH-positron has the opposite values of this
quantum-numbers, so SBH-electron and SBH-positron annihilate to a
radiation of 2-times the energy of the stimulated mass $\Delta m_{\rm n}$,
just as electron and positron do.
When two SBH-electrons meet, they will not merge to a double-massive
SBH-electron, as the two $\hbar/2$-spins would add to $\hbar$ or $0$,
but the double-massive SBH-electron must have a $\pm \hbar/2$-spin.

\section{Conclusion}

The Large Numbers Hypothesis was discovered by Dirac in 1937 in
which he found that cosmological quantities can
be related to particle quantities by simple scale relations of
which he thought could not just be random in nature.

Gravitation is a sort of energy and in consequence
itself is a source of gravitation.
Such a self-relating characteristic is a
typical element of any fractal structure.
Here a blueprint of a fractal Universe composed of black holes was
outlined by assuming that a gravitational wave
should not overrun its own event-horizon.
The model gives possible answers to still miraculous issues, such as the
entropy of black holes, the origin and integer value of the electric charge,
the spin, sizes and weights of elementary particles and the background
of Dirac's Large Numbers coincidences.

\end{document}